# Primordial Regular Black Holes: Thermodynamics and Dark Matter

**José Antonio de Freitas Pacheco**

Observatoire de la Côte d'Azur, Laboratoire Lagrange, 06304 Nice Cedex, France; pacheco@oca.eu; Tel.: +33-492-003-182

**Abstract:** The possibility that dark matter particles could be constituted by extreme regular primordial black holes is discussed. Extreme black holes have zero surface temperature, and are not subjected to the Hawking evaporation process. Assuming that the common horizon radius of these black holes is fixed by the minimum distance that is derived from the Riemann invariant computed from loop quantum gravity, the masses of these non-singular stable black holes are of the order of the Planck mass. However, if they are formed just after inflation, during reheating, their initial masses are about six orders of magnitude higher. After a short period of growth by the accretion of relativistic matter, they evaporate until reaching the extreme solution. Only a fraction of $3.8 \times 10^{-22}$ of relativistic matter is required to be converted into primordial black holes (PBHs) in order to explain the present abundance of dark matter particles.

## 1. Introduction

The detection of gravitational waves emitted during the merger of two black holes [1,2] represents a robust demonstration of the reality of these objects. Previously, the study of the motion of several individual stars around Sgr A*, a radio source located in the galactic center, led to the conclusion that the orbits of those stars are controlled by the gravitation of a "black" object having a mass of about $4 \times 10^6$ $M_\odot$ [3]. These observations strongly suggest the presence of a supermassive black hole in the center of the Milky Way, since no other adequate alternative for the nature of such a massive object was proposed up until now. Thus, the existence of "stellar" black holes with masses of few tens of the solar mass, or supermassive black holes with masses of six up to nine orders of magnitude the mass of the Sun, seems to be well established.

From a simple mathematical point of view, a black hole represents a region of the space–time causally disconnected from observers located at arbitrarily large distances. The surface separating both regions of the space–time is the event horizon that, in fact, is a one-way membrane, since observers inside the horizon can receive signals from outside, but the reverse is not true.

A massive object whose mass has been reduced to a "point singularity" of the space–time after it underwent gravitational collapse, represents an unpleasant physical situation because singularities denote points of the space–time where the classical theory breaks down. Einstein equations admit both past and future singularities hidden by an event horizon [4,5], but this theory in incomplete, because it ignores the presence of quantum effects. Such effects are expected to become significant in the high curvature regions existing near the singularity, which modify the space–time structure and make unreliable general relativity predictions.

Loop quantum gravity (LQG) has emerged in the past decades as a possible candidate for a quantum gravity theory, and investigations of the interior inside the event horizon based on LQG led to solutions free of the classical singularity. This is a consequence of the space–time continuum of general relativity being replaced by a discrete quantum geometry, which remains regular at the classical singularity. A complete treatment of the space–time of a black hole in LQG is still lacking, but different studies suggest that a singularity is not formed at the end of the gravitational collapse [6–8]. However, there are many uncertainties about the solution after the "crossing" of the classical singularity, or even about what replaces the black hole singularity. Some investigations [9] indicate that the structure of the solution allows the existence of a Cauchy horizon near *r* = 0. Thus, after the gravitational collapse, LQG solutions suggest that the interior inside of a black hole can be described by a singularity-free Reissner–Nordström space–time, including a Cauchy and an event horizon.

The LQG picture described above has similar analogs in general relativity, where regular metrics describing black holes have been proposed by Bardeen [10] and Hayward [11], among others. These solutions are non-singular, and both have an inner Cauchy horizon as well as an outer event horizon. Inside the Cauchy horizon, both geometries behave like a de Sitter space–time. It is interesting to mention that already in the sixties, Andrei Sakharov [12] had the intuition that at very high densities, the matter approaches a vacuum state with a finite density. Such a non-divergence implies that the local geometry should be described by a de Sitter metric.

In this paper, some aspects of these regular solutions will be reviewed, as well as, in particular, their associated thermodynamic properties. Both Bardeen and Hayward space–times are described by metrics characterized by free parameters that define the mass density distribution. Here, these scale parameters are fixed under the assumption that the radius of the Cauchy horizon is equivalent to the minimum distance derived from LQG. Under this condition, extremal black holes have masses of the order of the Planck mass. Then, based on thermodynamic principles, it will be shown that primordial regular black holes of about $10^6$ times the Planck mass can be formed at the end of the inflationary epoch, when the oscillations of the inflaton field are intense, and reheating occurs. These newly formed black holes have a short period of growth, and then they evaporate until reaching masses close to the extremal case. Such black hole remnants are possible candidates for dark matter particles. It will be shown that only a small fraction of relativistic matter ($3.8 \times 10^{-22}$) needs to collapse into black holes in order to explain the present dark matter abundance. The paper is organized as follows. In Section 2, the main properties of the Bardeen and the Hayward black holes are reviewed, while the thermodynamic properties of these objects are discussed in Section 3. Then, in Section 4, the formation of regular black holes in the early universe is considered, and finally, in Section 5, the main results are discussed.

## 2. Regular Black Holes

Since the original investigation of regular black holes by Bardeen [10], different studies have been addressed to the analysis of non-singular space–times [13–15]. In the case of static space–times, the considered general metric (in geometric units) is:

$$ds^2 = -f(r)dt^2 + f^{-1}(r)dr^2 + r^2 d\Omega^2 \tag{1}$$

where $f(r)$ is the lapse function. The zeros of the lapse function define the position of the event horizon and the inner Cauchy horizon, when it exists. In the case of the Bardeen geometry, the lapse function is defined by:

$$f(r) = 1 - \frac{2Mr^2}{(r^2 + g^2)^{3/2}} \tag{2}$$

In the above equation, *M* is the mass of the black hole, and *g* is a suitable scale. Ayon-Beato and Garcia [16] interpreted the scale parameter *g* as the monopole charge of a magnetic field in the context of a non-linear electrodynamic theory. Here, such a parameter is considered simply as a scale defining the mass distribution derived from Einstein equations, that is:

$$\rho(r) = \frac{3Mg^2}{4\pi(r^2 + g^2)^{5/2}} \tag{3}$$

Notice that when $r \gg g$, the lapse function reduces to the Schwarzschild geometry, while when $r \to 0$, the metric becomes essentially de Sitter, namely, regular at $r = 0$. Define now two dimensionless variables $x$ and $\gamma$ as:

$$x = \frac{r}{2M} \quad \text{and} \quad \gamma = \frac{g}{2M} \tag{4}$$

It worth mentioning that the quantities above are dimensionless in geometric units, but their real physical dimensions are always recovered when necessary. Studying the zeros of the lapse function, the existence or not of horizons depends on the following condition: if $\gamma < 2/\sqrt{27}$, two horizons exist, while if the inequality is not satisfied, there are no horizons. In the case of equality, the Cauchy and the event horizons coincide at the dimensionless coordinate $x_H = \sqrt{8/27}$. This corresponds to the case of an extreme Bardeen black hole.

The existence of horizons depends on the ratio between the scale parameter and the mass of the black hole. Is it possible to estimate the scale parameter $g$ in an independent way? A positive answer can be given within the following context: we will consider the extreme case for reasons that will be explained later, and we will consider that the common horizon radius $r_H$ is fixed by the minimum distance from the origin derived from the Riemann curvature invariant computed in terms of the volume operator in LQG [6]. Hence:

$$r_H = \sqrt{\frac{\pi}{2}} \ell_P \tag{5}$$

where $\ell_P$ is the Planck distance scale. From the relation above and the previous results, one obtains trivially that $g = \sqrt{(\pi/4)} \ell_P$, and that the mass of an extreme Bardeen black hole is:

$$M_* = \frac{\sqrt{27\pi}}{8} M_P \tag{6}$$

where $M_P$ is the Planck mass. Thus, our hypothesis concerning the scale parameter leads to a mass of the order of the Planck mass for an extreme Bardeen black hole. Using these results and Equation (2), the central density can be estimated (here the physical constants were recovered):

$$\rho_o = \frac{3\sqrt{27}}{4\pi^2} \frac{\hbar c}{\ell_P^4} \tag{7}$$

This result should be compared with the expected density derived from loop quantum cosmology (LQC) at the bounce (see, for instance, Craig [17]):

$$\rho_0 = \frac{\sqrt{3}}{32\pi^2 \gamma_I^3} \frac{\hbar c}{\ell_P^4} \tag{8}$$

In the equation above, $\gamma_I \simeq 0.2375$ is the so-called Barbero–Immirzi parameter. The numerical coefficient in Equation (7) is 0.395, while in Equation (8), it is about 0.409. Thus, under our assumptions, the expected central energy density for the extreme Bardeen black hole is comparable to the maximum density of the universe, which was attained at the instant of the bounce within the LQC scenario.

Similar computations can be performed in the case of the Hayward [11] metric that is characterized by the lapse function:

$$f(r) = 1 - \frac{2Mr^2}{(r^3 + 2ML^2)} \tag{9}$$

where $L$ is a scale parameter. In this case, the energy density distribution resulting from Einstein equations is:

$$\rho(r) = \frac{3M^2 L^2}{2\pi (r^3 + 2ML^2)^2} \tag{10}$$

Introducing as before the dimensionless quantities $x = r/2M$ and $\beta = L/2M$, the analysis of the roots of the lapse function indicate that there a critical value for the scale parameter $\beta_* = 2/\sqrt{27}$, corresponding to the critical coordinate $x_* = 2/3$. If $\beta$ is smaller than the critical value, two horizons exist while in the opposite situation, there is no black hole solution. The critical value defines the extreme case when both horizons coincide.

The scale parameter is fixed by assuming as before that the radial coordinate of the critical solution is equal to the minimum distance derived from LQG. In this case, one obtains:

$$L = \sqrt{\frac{\pi}{6}} \ell_P \tag{11}$$

From this result, the mass of the extreme Hayward black hole is:

$$M_* = \frac{3\sqrt{\pi}}{4} M_P \tag{12}$$

which is a value slightly greater than the Planck mass. The central energy density in this case is:

$$\rho_0 = \frac{9}{4\pi^2} \frac{\hbar c}{\ell_P^4} \tag{13}$$

which is smaller than that derived for a Bardeen black hole approximately by factor of two.

## 3. Thermodynamics of Regular Black Holes

An important breakthrough in the theory of black holes was the recognition that the laws of the mechanics governing the structure of these objects are analogous to the laws of thermodynamics, when the gravity at the horizon and the surface of the horizon are associated respectively to temperature and entropy [18,19]. Such an analogy was reinforced by the discovery by Hawking [20] that black holes can emit radiation as a grey body at the temperature defined by the horizon gravity. In fact, the emission spectrum, including particles other than photons, has a Planckian form only if the black hole is uncharged and non-rotating [21]. As a consequence of such a radiation, small black holes "evaporate", and primordial black holes with masses less than $3 \times 10^{14}$ g have already disappeared by now. It is worth mentioning that the Hawking radiation reinforces the connection between mechanic and thermodynamic laws, suggesting that the horizon surface should be interpreted as the physical entropy and the surface gravity as the physical temperature of the black hole.

The evaporation process raises some questions as, for instance: do black holes evaporate completely without leaving a remnant? Is the singularity suppressed at the end of the evaporation process [22,23]? String theory suggests a possible modification of the Heisenberg uncertainty principle, such as:

$$\Delta \vec{x} \Delta \vec{p} \approx \frac{\hbar}{2}\left(1 + \alpha^2 \ell_P^2 \frac{\Delta \vec{p}^2}{\hbar^2}\right) \quad (14)$$

where $\alpha$ is a constant of the order of the unity representing the string tension. Using the above relation, it is possible from first principles to estimate the associated black hole temperature at the horizon (see, for instance, Adler et al. [24]):

$$kT = \frac{Mc^2}{\pi \alpha^2}\left[1 - \sqrt{1 - \frac{\alpha^2 M_P^2}{4M^2}}\right] \quad (15)$$

For masses greater than the Planck mass, the usual result is recovered. However, Equation (15) indicates that during the evaporation process, the black hole mass reaches a minimum value $M_{min} = \alpha M_P / 2$; otherwise, the temperature becomes imaginary. In this case, the remnant of the evaporation process has a mass of the order of the Planck mass. However, according to Equation (15), such a remnant has a finite temperature, which is in fact a maximum, given by:

$$kT_* = \frac{M_P c^2}{2\alpha\pi} \quad (16)$$

This is an unpleasant physical situation, since the surface temperature of the remnant is not zero, and no radiation is allowed, since the mass cannot decrease. However, a thermally stable situation exists for extreme black holes, since the horizon temperature is zero. This is the case for an extreme Reissner–Nordström black hole, as well as for extreme Bardeen and Hayward black holes, as we shall see below.

The horizon temperature is given by the well-known relation:

$$T_H = \frac{f'(r_H)}{4\pi} \quad (17)$$

Using the lapse function corresponding to the Bardeen metric and the dimensionless variables defined previously, one obtains for the temperature:

$$T_H = \frac{1}{8\pi M}\left(1 - \frac{2\gamma^2}{x_H^2}\right) x_H^{-1/3} \quad (18)$$

Recalling that for the extreme case $\gamma_* = 2/\sqrt{27}$ and $x_* = \sqrt{8/27}$, it is trivial to verify from Equation (18) that the horizon temperature is zero. For non-extreme Bardeen black holes, the variation of the temperature as a function of the horizon radius is shown in Figure 1.

An inspection of Figure 1 reveals that for $(r_H / \ell_P) \gg 1$, the horizon temperature of the Bardeen black hole approaches the Schwarzschild behavior, as expected, decaying proportionally to the inverse of the horizon radius. For smaller masses, deviations between both temperatures become important when the horizon radius is of the order of $80\ell_P$. Contrary to the Schwarzschild case, the Bardeen black hole has a maximum temperature at the horizon radius $r_H \simeq 2.39\ell_P$. For still smaller masses, the temperature drops quite fast, reaching the zero value (extreme case) at the horizon radius $r_H = \sqrt{(\pi/2)}\,\ell_P$.

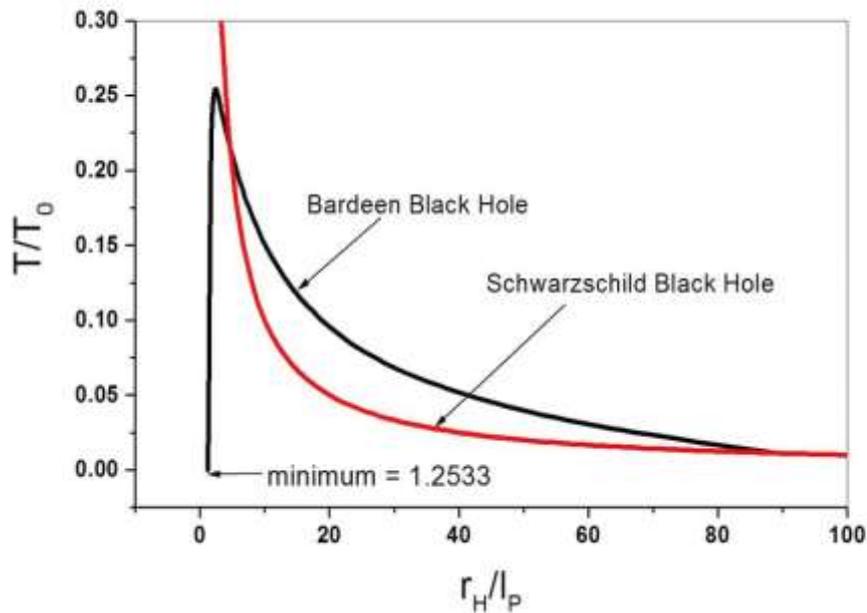

**Figure 1.** Normalized horizon temperature for Bardeen (black curve) and Schwarzschild (red curve) black holes as a function of the horizon radius in units of the Planck scale distance.

Similar calculations can be performed for the case of the Hayward metric, which permits obtaining the horizon temperature in terms of the dimensionless quantities defined previously, that is:

$$T_H = \frac{1}{8\pi M}\left(1 - \frac{3\beta^2}{x_H^2}\right) x_H^{-1} \qquad (19)$$

Since for an extreme Hayward black hole $x_* = 2/3$ and $\beta_* = 2/\sqrt{27}$, it is easy to verify that the temperature is zero, as expected. The temperature for Hayward black holes of different masses was computed numerically, and it is shown in Figure 2.

Notice that already for black holes with a horizon radius higher than $7\ell_P$ the temperatures for both black holes are practically indistinguishable. For lower masses, the Schwarzschild temperature diverges, while for the Hayward black hole, the temperature reaches a maximum near $r_H \simeq 2.17\ell_P$ and becomes zero again at $r_H = \sqrt{(\pi/2)}\,\ell_P$, which is the extreme case. This later value is equal to the precedent case, since, by assumption, the horizon radius of both black holes was assumed to be equal to the minimum distance derived from LQG.

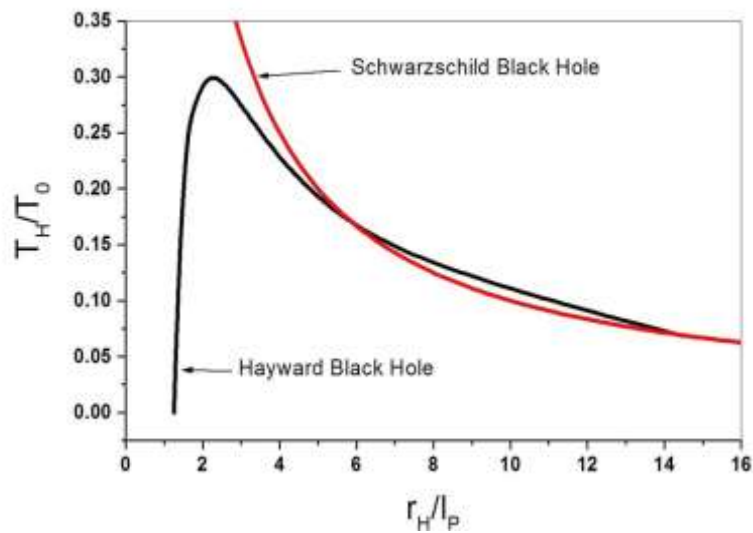

**Figure 2.** Normalized horizon temperature for Hayward (black curve) and Schwarzschild (red curve) black holes as a function of the horizon radius in units of the Planck scale distance.

It is interesting to evaluate the specific heat of these regular black hole solutions, which give some additional insight on their thermal properties. Define the quantity:

$$c_V = \frac{1}{k}\frac{\partial E}{\partial T} \qquad (20)$$

which represents the dimensionless specific heat. In order to evaluate the derivative above, it is necessary to specify a prescription for the energy (including the contribution of the gravitational field), which is an ill defined quantity in the general relativity theory. A compilation of energy-momentum complexes for different prescriptions was performed by Virbhadra [25], and here, the Einstein formulation was adopted. In this case, for the metric defined by Equation (1), the energy enclosed by a spherical surface of area A and radius $r = (A/4\pi)^{1/2}$ is:

$$E = \frac{c^4 r}{2G}\left[1 - f(r)\right] \qquad (21)$$

If the considered surface is the horizon, $f(r_H) = 0$ and the energy is proportional to the horizon radius. Consequently, in the case of the Schwarzschild metric, the energy is simply given by the mass of the black hole, that is, $E = Mc^2$. Since for a Schwarzschild black hole, the horizon temperature is inversely proportional to the mass, from Equation (20), one obtains a negative specific heat, which is a well-known result. This means that if the black hole absorbs energy, its temperature decreases. However, the situation is not exactly the same for the regular black holes here discussed. Firstly, since the energy enclosing the horizon depends directly on the horizon radius, its trivial to show that:

$$E = xMc^2 \qquad (22)$$

where the variable $x$ was defined by Equation (4), and its value is derived from the zeros of the lapse function. For an extremal Hayward black hole, $x = 2/3$, while for a Bardeen black hole, $x = \sqrt{8/27}$. This means that these extremal black holes have energies smaller than a Schwarzschild black hole of the same mass. For larger masses $x \to 1$, all of the black holes of a given mass have the same energy inside the horizon. Secondly, since the energy and the temperature are functions of the horizon radius, the specific heat can be computed as:

$$c_V = \frac{1}{k} \frac{\partial E}{\partial r_H} \frac{\partial r_H}{\partial T} \qquad (23)$$

Computing the derivatives using the expressions for the temperature derived above, for the Hayward regular black hole, one obtains:

$$c_V = \frac{2\pi r_H^2}{k \ell_P^2} \left( \frac{9L^2}{r_H^2} - 1 \right)^{-1} \qquad (24)$$

This result indicates that the specific heat for Hayward black holes is positive if the horizon radius is in the range $\sqrt{3}L$ (extremal black hole) and 3L. The upper limit corresponds to the temperature maximum and the specific heat diverges here. Beyond this critical value, the specific heat becomes negative. This behavior is consistent with the trend observed in the horizon temperature curve (see figure 2). The extremal case has T=0, and the temperature increases as the mass increases, since the specific heat is positive. The maximum temperature occurs for $r_H = 3L = \sqrt{3\pi/2}\,\ell_P \approx 2.17\,\ell_P$, as mentioned previously. Once the specific heat becomes negative, the temperature decreases as the mass of the black hole increases, following the behavior observed for Schwarzschild black holes. A similar behavior occurs for the Bardeen case.

It is interesting to compute also the emission rate (luminosity) due to the Hawking process for these regular black holes. Assuming that the horizon radiates like a black body, the luminosity of the Bardeen black hole is given by:

$$L_B = K_B \gamma^2 x^{2/3} \left[ 1 - \frac{2\gamma^2}{x^2} \right]^4 \qquad (25)$$

where the quantities $x$ and $\gamma$ have the same meaning as before, and are derived by computing the zeros of the lapse function. The constant $K_B$ is defined as:

$$K_B = \frac{\sigma_s}{8\pi^4 \ell_P^2} \left( \frac{\hbar c}{k} \right)^4 \qquad (26)$$

where $\sigma_s$ is the Stephan radiation constant and $k$ is the Boltzmann constant. Similar computations can be performed for the Hayward metric, and one obtains:

$$L_H = K_H \beta^2 x^{-2} \left[ 1 - \frac{3\beta^2}{x} \right]^4 \qquad (27)$$

The constant in the equation above satisfies $K_H = 3K_B/4$, and the dimensionless quantities are again as before. Figure 3 shows a plot of the luminosities normalized either to $K_B$ or $K_H$ as a function of the horizon radius.

A simple inspection of Figure 3 indicates that the luminosity for both regular black holes does not diverge as in the Schwarzschild case, since they are zero for the extreme case, when $r_* = \sqrt{(\pi/2)}\,\ell_P$. In both cases, the luminosity reaches a maximum respectively at $r_H \approx 2.8\,\ell_P$ for the Hayward black hole, and at $r_H \approx 3.6\,\ell_P$ for the Bardeen black hole.

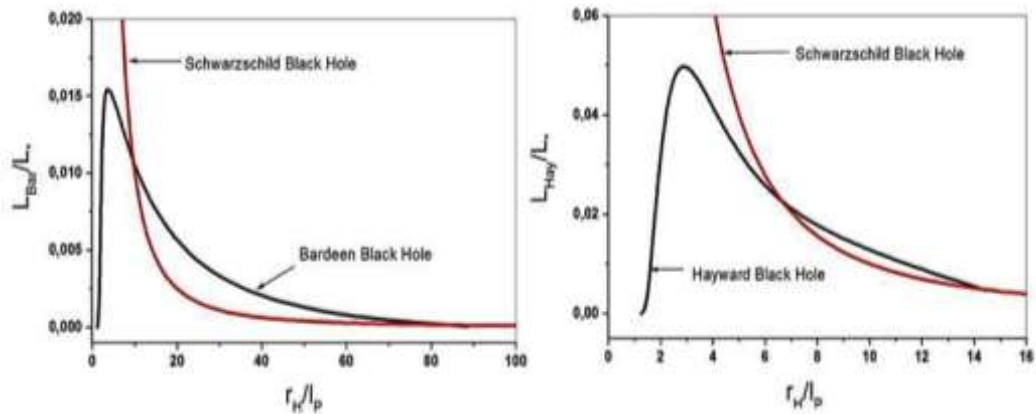

**Figure 3.** Normalized luminosities for the Bardeen (**left** panel) black hole and the Hayward (**right** panel) black hole as a function of the horizon radius in units of the Planck distance scale.

## 4. Primordial Regular Black Holes

In the previous sections, we have seen that the Bardeen and the Hayward regular black holes have a geometric structure similar to that derived from investigations of the gravitational collapse based on LQG; in other words, a space–time including two horizons without any singularity. The extreme case has zero surface temperature, and it is thermally stable if the black hole is isolated, which is in agreement with the detailed investigations performed in the case of the Reissner–Nordström metric [26,27]. Under the assumption that the horizon radius of these extreme regular black holes are equal to the minimal distance derived from LQG, it is possible to conclude that their masses are comparable to the Planck mass. These objects could be possible candidates to be identified with dark matter particles if they were produced in the early universe. This possibility was already suggested by MacGibbon [28] in the late eighties. He postulated the existence of black holes with Planck masses, which would be relics of the Hawking process.

The formation of black holes in the early universe was already considered either by Zeldovich and Novikov [29], or by Hawking [30]. These black holes are expected to be formed by the gravitational collapse of primordial density fluctuations in the radiation-dominated phase of the early universe. In order to collapse against matter pressure, the collapsing region must be larger than the Jeans length at maximum expansion. Moreover, the condition that the gravitational radius should be smaller than the particle horizon fixes the maximum mass of the black hole that can be formed in a given instant of time. Two aspects play a central role in the formation of primordial black holes (PBHs): first, for each horizon-sized region, there exists a critical threshold density contrast $\delta_c$, above which the collapse occurs. Comparing the Jeans and the horizon lengths at the time when the collapsing region breaks away from the Hubble expansion, one finds that the critical density contrast must be of the order of the unity. The second key assumption concerns the final mass of the black hole, which is commonly supposed to be approximately close to the horizon mass at the epoch of formation.

Investigations using a variety of initial density perturbation profiles, based on self-similarity and scaling led to a relation between the PBH mass and the horizon-scale of the form [31,32]:

$$M = KM_H(\delta - \delta_c)^\eta \quad (28)$$

where $M_H$ is the horizon mass scale, and $K$ and $\eta$ are constants. Since the PBH mass goes to zero as the density contrast is close to $\delta_c$, the existence of critical phenomena suggests the possibility that masses at formation could be much smaller than the horizon scale. Recent studies indicate that PBHs with a broad mass spectrum can be formed in the high peaks of the co-moving curvature power spectrum resulting from single field inflation [33]. However, the converted mass fraction into black holes is sensitive to possible non-Gaussianities in the amplitude distribution of such large and rare density fluctuations [34].

Astronomical observations can put severe constraints on the mass of PBHs that are able to explain the observed cosmological dark matter abundance. Data on extragalactic γ-rays, the femtolensing of γ-ray bursts, white-dwarf explosions, neutron-star captures, and quasar microlensing have been reviewed by Kühnel & Freese [35], and no severe limits exist either for PBH masses less than $10^{-17}$ $M_\odot$ or higher than 10 $M_\odot$. Hence, present astronomical data do not impose any constraint on the existence of Planck mass black holes, and on the interpretation of these objects as dark matter particles. However, an opposite direction has been taken by some authors, who have suggested that dark matter candidates are more massive black holes either with masses of about 30 $M_\odot$ [36] or in the range of $10$–$10^5$ $M_\odot$ [37].

*Lower Limits for PHBs Masses from Thermodynamics*

As we have seen, PBH masses cannot be larger than the horizon scale. On the low side of the mass spectrum, Equation (28) suggests that PBH with very small masses can be formed. We will assume that these PBHs are formed just after inflation during reheating. At end of the inflationary period, the inflaton field is subjected to strong oscillations and decay. Those oscillations can be the origin of density fluctuations that satisfy the conditions fixed by Equation (28), and thereby forming black holes. However, the PBH masses cannot be arbitrarily small, and limits are fixed by thermodynamics. Consider a small spherical perturbation with a co-moving volume V. The entropy of such a perturbation if constituted by relativistic matter is:

$$S_m = \frac{4\varepsilon V}{3kT} \qquad (29)$$

Let $\alpha < 1$ be the efficiency of the gravitational collapse. In this case, the mass of the resulting black hole is $M = \alpha(\varepsilon V)/c^2$. Since the entropy of the collapsing matter cannot be larger than that of the resulting black hole, the following condition must be satisfied:

$$\frac{4Mc^2}{3\alpha kT} < \frac{\pi r_H^2}{\ell_P^2} \qquad (30)$$

Using the dimensionless quantities defined in the previous section, in the case of the Bardeen space–time, the following condition relates the matter temperature *T* with the scale parameter *γ* and the black hole horizon $x_H$:

$$kT \geq \frac{4}{3\alpha\sqrt{\pi^3}} \frac{\gamma}{x_H^2} M_P c^2 \qquad (31)$$

For numerical purposes, the collapse efficiency is taken as $\alpha = 0.5$, and the reheating temperature is taken as $10^{12}$ GeV. Recalling that $\gamma$ and $x_H$ are connected by the equation $f(x,\gamma) = 0$, the numerical solution of these equations gives $\gamma \approx 1.7 \times 10^{-7}$. The associated black hole mass is:

$$\frac{M}{M_P} = \frac{\sqrt{\pi}}{4\gamma} \approx 2.6 \times 10^6 \qquad (32)$$

Similarly, for the Hayward case, the following condition must be satisfied:

$$kT \geq \frac{4\beta}{\alpha\sqrt{3\pi^3}} \frac{M_P c^2}{x_H^2} \qquad (33)$$

Assuming the same conditions as before, one obtains $\beta \approx 10^{-7}$, and a black hole mass:

$$\frac{M}{M_P} = \sqrt{\frac{\pi}{12}} \frac{1}{\beta} \approx 5 \times 10^6 \quad (34)$$

These results indicate that PBHs formed at reheating have masses that are a few million times higher than the masses of the extreme case. Hence, a natural question arises: do these PBHs evaporate until the stable state is attained, or do they grow by accretion of relativistic matter? In order to answer this question, the evaporation and the accretion timescales must be compared.

The evaporation timescale can be estimated from the relations for the Hawking luminosity derived previously, since $t_{evap} = Mc^2/L$. At formation, the Hawking luminosity of a Bardeen black hole is $L = 1.25 \times 10^{42}$ erg/s, implying an evaporation timescale of $4.0 \times 0^{-20} s$. A similar calculation for a Hayward black hole leads to a timescale of $1.7 \times 10^{-19} s$. On the other side, the accretion timescale can be defined as $t_{acc} = M/(dM/dt)$. In order to estimate the accretion rate, we have followed the same procedure as that by de Freitas Pacheco [38] concerning the accretion of relativistic matter by a Reissner–Nordström black hole. Since a detailed analysis of the accretion flow is beyond the purposes of this paper, only the main points are recalled here. The position of the critical point and of the radial velocity there can be generalized for a metric defined by Equation (1), and are given respectively by:

$$r_c = \frac{4 f(r_c)}{f'(r_c)} \frac{V^2}{(1-V^2)} \quad (35)$$

where $f(r)$ and $f'(r)$ are respectively the lapse function and its derivative with respect to the radial coordinate taken at the critical point. The generalized sound velocity is defined by:

$$V^2 = \frac{d \lg(P+\varepsilon)}{d \lg n} - 1 \quad (36)$$

In the case of a relativistic fluid $(P+\varepsilon) \propto n^{4/3}$, and one obtains trivially that $V = 1/\sqrt{3}$. The radial component of the four-velocity at the critical point is given:

$$u_c^2 = \frac{V^2}{(1-V^2)} f(r_c) \quad (37)$$

A numerical solution of these equations for the case of the Hayward metric is shown in Figure 4.

These calculations indicate significant deviations with respect to the Schwarzschild space–time for masses close to the Planck value. For larger masses, the critical radius and the radial velocity at this point approach asymptotically the Schwarzschild results, that is, $r_c \to 3r_g/2$ and $u_c \to 1/\sqrt{6}$. Notice that the crossing of the critical point occurs always subsonically.

Once the critical radius and the radial velocity at this point were computed, the accretion rate can be estimated from the equation:

$$\frac{dM}{dt} = \frac{4\pi}{c} r_c^2 T_t^r \quad (38)$$

where $T_i^k$ is the stress-energy tensor of the accreting fluid evaluated at the critical point. For the Hayward black hole formed at reheating, the estimated accretion rate is $9.1 \times 10^{26}$ g.s$^{-1}$, which corresponds to an accretion timescale of $1.2 \times 10^{-25} s^{-1}$. This value is six orders of magnitude smaller than the evaporation timescale derived above, indicating that the newly formed black hole will grow. The scenario is the same for the Bardeen case. However, the accretion rate either for the Hayward or Bardeen black holes depends on the energy density of the cosmic relativistic matter, which varies with the temperature as $\varepsilon \propto T^4$. Due to the fast expansion of the universe, the temperature decreases with a timescale $t_{col} = T/|dT/dt| = 1/H$. At reheating, this is about $4.7 \times 10^{-31} s$, which is several orders of magnitude smaller than the two other timescales. This means that these PBHs have initially a very short phase of growth, and then, the evaporation process dominates until the extreme situation is reached.

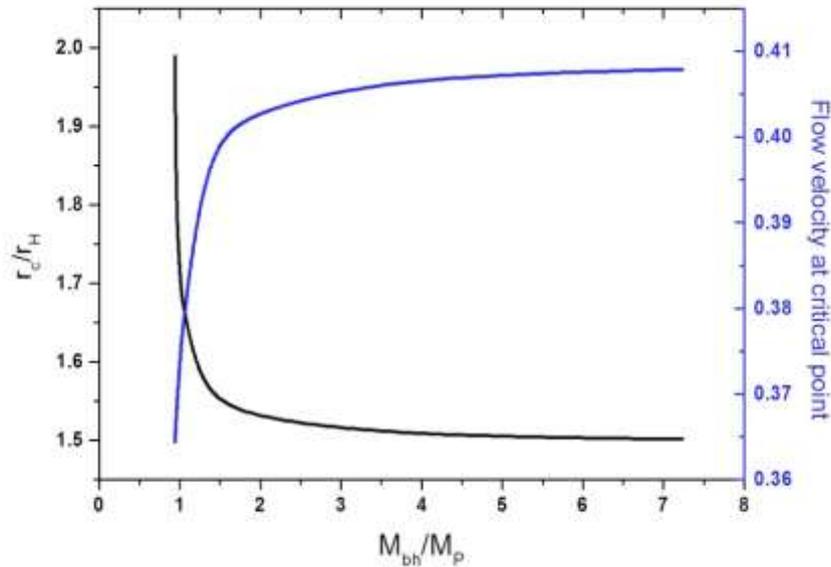

**Figure 4.** Variation of the critical radius in units of the gravitational radius as a function of the Hayward black hole mass in units of the Planck mass (black curve). The radial component of the four-velocity of the flow is also shown as a function of the black hole mass (blue curve).

If the density fluctuations giving origin to PBHs are assumed to be spherically symmetric with a Gaussian distribution and are scale invariant, the mass spectrum of the formed black holes is of the form [39]:

$$\frac{dN}{dM} \propto M^{-5/2} \tag{39}$$

This implies that the average mass of PBHs is $<M> = 3 M_{min}$, where $M_{min}$ is the minimum mass estimated previously on the basis of thermodynamic arguments. Consequently, most of the formed PBHs have masses approximately of the order of the minimum value, which will evolve according to the picture above.

Finally, it is useful to estimate the energy fraction of PBHs formed at reheating required to explain the present dark matter abundance. A simple calculation gives the ratio between the PBH energy density and the relativistic matter energy density at formation:

$$\frac{\rho_{reh}}{\varepsilon_{reh}} = \frac{\Omega_{dm}}{\Omega_\gamma} \left(\frac{g_0}{g_{reh}}\right)^{1/3} \left(\frac{T_0}{T_{reh}}\right) \tag{40}$$

where $\Omega_{dm}$ and $\Omega_\gamma$ are the present density parameters of dark matter and radiation, the $g_i$s are the number of degrees of freedom at the present epoch and at reheating, and the Ts are the temperatures at the same considered cosmological times. Numerically, Equation (40) gives the initial fraction between dark matter and radiation $3.8 \times 10^{-22}$. Hence, only a very small fraction of the relativistic matter needs to undergo the gravitational collapse in order to explain the observed amount of dark matter. Notice that such a small value is a consequence of the different dilution factors for non-relativistic and relativistic matter as the universe expands as well as the huge mass of the particles (~$10^{19}$ GeV/c²), implying presently that only a low-particle density is required to explain the observations.

## 5. Discussions

Up until now, there has been no direct or indirect evidence for dark matter, which would be respectively the consequence of collisions with baryons in the laboratory, or the annihilation resulting from collisions between particles and antiparticles of dark matter. Gravitational effects remain the only source of inference concerning the existence of such a form of exotic matter. In particular, primordial density fluctuations in a universe constituted only by baryons, whose amount is fixed by primordial nucleosynthesis and the density of relic photons, will never reach the non-linear regime in timescales of 13–14 Gyr. Consequently, galaxies could not presently exist [40,41]. However, among particles issued from the Standard Model (SM), there are no relic candidates with the required DM abundance that is able to explain the observations. Neutrinos are generally excluded because they are not massive enough, and decouple relativistically from the cosmic plasma, constituting a model for "hot" dark matter, which has problems with the matter power spectrum at scales smaller than $10^{14}-10^{15} M_\odot$. As a result of these difficulties, modifications of the Standard Model have been proposed, in particular a minimal supersymmetric extension (MSSM). In this model, the neutralino, the lightest supersymmetric particle, is the "preferred" candidate [42]. However, from the experimental side, there are many difficulties with this theory, since up to now, no signal of supersymmetry has been seen in experiments related to the decay of B mesons, which do not indicate the presence of "exotic" particles such as charginos and/or neutralinos [43,44].

These tensions between astronomical and physical data led to alternative proposals such as modifications of general relativity theory [45], dark matter particles having masses of about few MeV/c² [46], or on the contrary, having masses around few TeV/c², resulting from the SO(10) breaking [47]. In the present work, the possibility that primordial regular black holes could be identified with dark matter particles was investigated. This possibility is not new, and past studies always had difficulties with the existence or not of remnants left by the evaporation process. Investigations of the gravitational collapse based on LQG suggest the appearance of a non-singular space–time with a Reissner–Nordström-like metric or, in other words, including two horizons. This behavior is well reproduced by regular black holes, whose geometry is described either by the Bardeen or the Hayward metric.

The Bardeen and the Hayward space–times approach the de Sitter solution in the region inside the Cauchy (or inner) horizon, implying that the equation of state of matter is similar to that of the vacuum ($P = -\rho$). In both metrics, there is a critical solution in which the two horizons coincide, representing the case of an extreme black hole. Extreme black holes have zero surface temperature, and consequently, no Hawking emission is present. These objects are thermally stable, and can be imagined as being in its "ground" state. The basic assumption of the present investigation is to consider that the horizon radius of these extreme black holes is equal to the minimum distance derived from the Riemann invariant computed from LQG. Under this hypothesis, the extreme black hole mass either in the Bardeen or in the Hayward case is of the order of Planck mass. Consequently, our hypothetical dark matter particle does not evaporate, and does not hide any space–time singularity. It is worth mentioning the work by Dymnikova and Khoplov [48], who considered regular black holes whose space–time is asymptotically de Sitter instead of Minkowski. These regular black holes have three horizons: the inner or Cauchy, the event, and the cosmological. According to these authors, regular PBHs with de Sitter interiors are formed when the collapse of a primordial fluctuation does not lead to a central singularity, stopping at a given very high density with a vacuum equation of state, as conjectured by Sakharov [12] many years ago. In this case, the de Sitter vacuum is formed with an energy density corresponding to that of the GUT symmetry restoration scale. Under these conditions, the Dymnikova–Khoplov regular black hole has a mass of about $2.7 \times 10^7 M_P$.

If our considered regular black holes are formed at the end of inflation, when the strong oscillations and decay of the inflaton field occur, their minimum mass is fixed by the condition that the entropy of the collapsing matter must be lower than the resulting black hole entropy. If the reheating temperature is $10^{12}$ GeV, the minimum masses are respectively $2.6 \times 10^6 M_P$ for the Bardeen solution, and $5.0 \times 10^6 M_P$ for the Hayward case. Recently, a similar scenario has been investigated [49], in which black hole formation occurs during the oscillatory phase after inflation in conditions of slow reheating. The authors have estimated that the minimum black hole mass at formation is $M_{\min} = 4\pi M_P (M_P / H_*) \sim 10^6 M_P$, since the expansion rate during inflation derived from Planck 2015 is $H_* \approx 10^{14} GeV$. Notice that this value compares quite well with our own estimates based on thermodynamic arguments.

As we have shown above, the newly formed black holes have initially a short phase of growth, followed by an evaporation phase, in which they lose mass until the stable extreme condition is reached. If the initial mass spectrum has the form given by Equation (39), most of the PBHs have masses that are few times the minimum value, and therefore, all of these objects will follow the same evolutionary path leading to the same end point. In conclusion, dark matter particles could be constituted by regular PBHs with masses around $10^{19}$ GeV/$c^2$, and only a very small fraction of the relativistic matter at reheating is needed to be converted into PBHs in order to explain the observed dark matter abundance, in agreement with the estimates made by Carr et al. [49].

**Conflicts of Interest:** The authors declare no conflict of interest.